\documentclass{article}
\usepackage{graphicx}
\usepackage{hyperref}
\usepackage{amsmath, amssymb}
\usepackage{physics}
\usepackage{cleveref}
\usepackage{bm} % bold math
\usepackage{xcolor}
\usepackage[noblocks]{authblk}

\usepackage{orcidlink}
\usepackage[sorting=none]{biblatex}
\providecommand{\keywords}[1]{{\small\textbf{\textit{Keywords---}}#1}} 

\newcommand{\abrc}[1]{\left\langle#1\right\rangle}
\newcommand{\brc}[1]{\left(#1\right)}
\newcommand{\Brc}[1]{\left[#1\right]}
\newcommand{\cbrc}[1]{\left\{#1\right\}}
\newcommand{\vbrc}[1]{\left|#1\right|}

\newcommand{\package}[1]{\textsc{#1}}

\begin{document}

\title{Performance Benchmarking: Software for the Density Matrix Renormalization Group}

\author[1,\footnote{Corresponding author: pers@cs.umu.se}]{Per Sehlstedt\orcidlink{0009-0008-9361-9377}}
\author[1]{Paolo Bientinesi\orcidlink{0000-0002-4972-7097}}
\author[1]{Lars Karlsson\orcidlink{0000-0002-4675-7434}}

\affil{Department of Computing Science, Umeå University, Umeå, 901 87, Sweden}

\maketitle

\begin{abstract}
    The performance of scientific software often determines the scale of problems that can be solved in practice. As multiple implementations of the same algorithm emerge, systematic evaluation is needed to compare their strengths and limitations. The density matrix renormalization group (DMRG) algorithm, widely used to study quantum systems, has over 50 software implementations. These implementations vary in multiple aspects that can strongly affect performance. However, despite the need, performance evaluations of these implementations are scarce and lack a consistent standard; many existing evaluations are either too incomplete to enable meaningful comparisons or focus on objectives other than direct performance comparisons, thereby limiting understanding of how the implementations compare. Here, we present a performance-oriented benchmarking framework to facilitate meaningful comparisons of DMRG implementations, and we apply it to quantify the performance of eight implementations, highlighting similarities and differences among them. Furthermore, we examine multiple parameter settings, optimization strategies, and implementation-specific features to demonstrate how parameter configuration can affect performance and how systematic evaluation can reveal non-obvious trade-offs. The results show significant performance differences, up to two orders of magnitude in some cases---not only between different implementations when aligning parameters, but also within the same implementation when comparing different parameter configurations. Hence, our results demonstrate the significant value and insight that can be gained from conducting rigorous performance evaluations. Using our results and framework as a starting point, more rigorous benchmarking will ultimately help users and developers make informed decisions and support future development efforts to build better, more efficient software. 
\end{abstract}


\section{Introduction}
\label{sec: Introduction}

Software excellence is characterized by multiple qualities; however, for the computational scientist, performance is a central concern, as it often determines the feasible scale of computation. Yet, when attempting to identify the most effective solution, the landscape can be crowded with competing implementations whose surface characteristics offer limited insight into practical behavior. Thus, meaningful judgment requires systematic evaluation.

These considerations are particularly acute for packages that implement the density matrix renormalization group (DMRG) algorithm \cite{White1992, White1993}, as their performance is sensitive to implementation details and parameter configuration. As the DMRG field has matured, more than 50 software implementations have emerged, each offering different trade-offs in functionality, ease of use, extensibility, and supported features \cite{Sehlstedt2026}. For new users in particular, navigating this landscape can be difficult: documentation quality varies, advertised capabilities are not always directly comparable, and practical guidance on which package to choose for a specific use case is often limited. Rigorous performance benchmarking would alleviate some of the difficulty by enabling meaningful comparisons across implementations. Additionally, it would help track progression and guide future development.

Currently, performance evaluations of DMRG implementations remain scarce and lack a consistent standard, with few cross-comparisons. Moreover, many existing evaluations are too incomplete to enable meaningful performance comparisons, as they omit critical context on the interplay between computational cost and solution quality. For instance, reporting a fast execution time in isolation provides limited value, since the speed may come at the expense of convergence or accuracy. Thus, without a clear connection between these aspects, our ability to assess how each package truly performs relative to the others is hindered.

In this paper, we pursue three aims: (i) In order to facilitate meaningful performance evaluations and comparisons of DMRG implementations, we present a performance-oriented benchmarking framework that focuses on the interplay between computational cost and solution quality; (ii) in order to highlight similarities and differences in performance across available implementations, we apply the framework and quantify the performance of eight implementations; (iii) in order to demonstrate how parameter configuration can affect performance and how systematic evaluation can reveal non-obvious trade-offs, we examine multiple parameter settings, optimization strategies, and implementation-specific features.
As such, we aim to establish a foundation for systematically evaluating tensor network methods such as DMRG, benefiting both users and developers by enabling more insightful performance assessments and better-informed decisions, and by supporting future development efforts. 

The remainder of this paper is structured as follows: We outline the DMRG algorithm and many of its tunable parameters and optimization strategies in \Cref{sec: Theory}; we review related work and describe our benchmarking framework in \Cref{sec: Performance benchmarking}, and we explain how we apply it in our experiments in \Cref{sec: Experimental setup}; we present the results of the experiments in \Cref{sec: Results} and provide the final discussions and conclusion in \Cref{sec: Discussion}.

\section{Theory}
\label{sec: Theory}

\subsection{The DMRG algorithm}

The DMRG algorithm~\cite{White1992, White1993} is today typically framed as a variational optimization seeking to minimize the expectation value $E$ of a given Hamiltonian $\widehat{H}$ over the set of states spanned by a variational ansatz $\ket{\psi}$:
\begin{equation}
    \arg\min_{\ket{\psi}} E(\ket{\psi}) = \arg\min_{\ket{\psi}} \frac{\ev{\widehat{H}}{\psi}}{\braket{\psi}}.
\end{equation}
The variational ansatz is usually chosen from a tensor network state family, as they have become recognized as a natural language for describing quantum states of matter \cite{Schollwck2011, Orus2014}.

The reason tensor network states constitute a natural choice for $\ket{\psi}$ is that they can provide efficient parameterizations by capturing the relevant entanglement properties of the systems under consideration.
Over the years, numerous tensor network families have been developed, each associated with distinct geometrical structures suited to different physical settings and classes of quantum states \cite{Cirac2009, Evenbly2011, Orus2019}.
A viable option is provided by tree tensor networks (TTN) \cite{Shi2006, Nakatani2013}; however, by far the most common choice within DMRG is the matrix product state (MPS) \cite{Schollwck2011}---also known as the tensor-train decomposition~\cite{Oseledets2011}.

A general MPS with open boundary conditions takes the form
\begin{equation}
    \label{eq: general MPS}
    \ket{\psi_{\mathrm{MPS}}} = \sum_{\bm{\sigma}} \brc{\prod^L_{k=1} M^{\sigma_k}} \ket{\bm{\sigma}}, \qquad \ket{\bm{\sigma}} = \bigotimes^L_{k=1} \ket{\sigma_k},
\end{equation}
where $L$ is the number of sites, $\cbrc{\ket{\sigma_k}}$ is the basis for the local Hilbert space at site $k$ with physical index $\sigma_k$ labeling the basis, and $\bm{\sigma}$ being the multi-index of all local state indices. 
Moreover, $M^{\sigma_k} \in \mathbb{C}^{D_k \times D_{k+1}}$ are local matrices, where $D_{k+1}$ denotes the bond dimension associated with the virtual link between sites $k$ and $k+1$, with $D_1 = D_{L+1} = 1$. The entries of these matrices constitute the variational parameters, with the bond dimension determining the amount of entanglement that can be represented \cite{Schollwck2011}.

Due to the matrix product structure in \cref{eq: general MPS}, the MPS is invariant under the insertion of an invertible matrix and its inverse between two adjacent sites, reflecting a gauge freedom.
This gauge freedom can be fixed by imposing left- and right-orthonormality of the matrices to the left and right of a chosen site, respectively, yielding a canonical form.
The canonical form plays a central role in DMRG, as it improves numerical stability and performance \cite{Schollwck2011}. 

In analogy with an MPS, the Hamiltonian will typically be represented as a matrix product operator (MPO), taking the form
\begin{equation}
    \widehat{H} = \sum_{\bm{\sigma}, \bm{\sigma'}} \brc{\prod^L_{k=1} W^{\sigma_k\sigma'_k}} \ket{\bm{\sigma}} \bra{\bm{\sigma'}},
\end{equation}
where the $W^{\sigma_k\sigma'_k} \in \mathbb{C}^{D_k \times D_{k+1}}$ matrices are analogous to the $M^{\sigma_k}$ matrices, but carry two physical indices instead of one, corresponding to the input and output states of the local Hilbert space \cite{Schollwck2011}.

DMRG minimizes $E$ via an iterative sweeping procedure. In this procedure, one or two adjacent local MPS tensors are optimized at a time while the others are held fixed, reducing the global optimization problem to a sequence of local optimizations. Using a canonical MPS form ensures that each local optimization remains well-conditioned throughout the procedure. At each sweep step, the effective eigenvalue problem is solved with an iterative sparse-matrix eigensolver, such as the Lanczos or Davidson methods. These projection-based solvers are preferred because they enable matrix-free implementations and efficiently target the desired extremal eigenpairs \cite{Schollwck2011}.

Each local optimization temporarily breaks the MPS's canonical form, which must be restored before proceeding to maintain the algorithm's efficiency. During this restoration, truncation is often necessary in the 2-site DMRG algorithm to control the enlarged bond dimension between sites, but the restoration can also be used to naturally allow for adaptive adjustment. Truncation is achieved by retaining the most significant states, where significance may be defined by a fixed maximum bond dimension, a threshold on the discarded weight, a combination of both, or other criteria. It is typically determined via a singular value decomposition, yielding an optimal low-rank approximation in the variational sense \cite{Schollwck2011}.

Symmetries in quantum systems give rise to conserved quantities, which often can be leveraged in DMRG to reduce computational effort.
If $\widehat{H}$ commutes with an operator $\widehat{Q}$,
\begin{equation}
    [\widehat{H}, \widehat{Q}] = 0,
\end{equation}
then $\widehat{Q}$ is a conserved charged, and the eigenstates of $\widehat{H}$ can be chosen simultaneously as eigenstates of $\widehat{Q}$.
The eigenvalues $q$ of $\widehat{Q}$ are thus good quantum numbers, and can be used to decompose the Hilbert space into independent sectors, making $\widehat{H}$ block-diagonal with respect to this decomposition.
Since the ground state lies within a definite symmetry sector, DMRG can restrict its variational search to the corresponding block, ensuring that the conserved quantity is preserved exactly throughout the calculation and avoiding unnecessary exploration of other sectors  \cite{Schollwck2011}.

Incorporating abelian symmetries---such as a $\mathrm{U}(1)$ symmetry associated with total magnetization or particle number conservation---remains simple to implement due to their commutativity, and typically provides notable efficiency gains.
Leveraging non-abelian symmetries---such as $\mathrm{SU}(2)$ total spin conservation---comes at a significantly higher implementation cost due to the complexity of managing non-commutative group representations, but can capture a richer internal structure and yield even greater efficiency gains~\cite{McCulloch2002, Weichselbaum2012}.

\subsection{Tunable parameters and optimization strategies}

Although the DMRG algorithm admits a compact high-level description, its practical performance depends on numerous implementation details and parameter configuration. Sensible default choices can provide robust performance in many cases, but no single set of defaults is optimal across all models and regimes. Consequently, it is important not only that users have access to these parameters but also that they are aware of different optimization strategies and which aspects of the algorithm can be tuned, so that they can make informed adjustments to adapt the method to the specific structure and numerical challenges of their problem. 

Here, we go through 12 different aspects.

\paragraph{Variational ansatz}
The most common choice for the variational ansatz within DMRG is the MPS; however, there are other options, such as the TTN, whose hierarchical structure can be beneficial in certain situations.

\paragraph{Update mode}
2-site DMRG offers greater robustness by naturally allowing the bond dimension to grow dynamically, at the expense of increased computational cost. Traditional 1-site DMRG, by contrast, is computationally cheaper per step but may require additional stabilization mechanisms to avoid local minima. Because of these trade-offs, the optimal approach can vary across different stages of the optimization. 

\paragraph{Bond dimension}
The bond dimension is one of the most influential parameters, as it determines the achievable accuracy. Beyond specifying a maximum bond dimension, users may enforce a minimum bond dimension or prescribe a gradual increase in it over time. Such strategies can improve convergence, particularly in early sweeps, and help avoid premature loss of relevant states during truncation~\cite{Stoudenmire2012}.  
\paragraph{Eigensolver}
The local eigensolver is a critical component, as its routine is typically the most time-consuming part of the algorithm. 
At a high level, the choice of method---such as Lanczos, Davidson, or related variants---already influences convergence behavior and computational cost. At a more fine-grained level, solver-specific settings, including convergence tolerances, Krylov subspace dimensions, cutoffs, restart strategies, preconditioners, and orthogonalization schemes, can further affect performance.

\paragraph{Truncation criterion}
The choice of truncation criterion---whether based on a fixed bond dimension, a discarded weight tolerance, or related entropy-based measures---can influence the trade-off between accuracy and efficiency. As with decisions about bond dimension, different stages of the optimization can benefit from different criteria; for example, progressively decreasing the cutoff can keep early sweeps efficient while ensuring that the final MPS achieves high accuracy.

\paragraph{Enhancements}
Stabilization and acceleration mechanisms have been introduced since the original formulation of DMRG. These include density matrix perturbation \cite{White2005}, subspace expansion \cite{Hubig2015}, and controlled bond expansion \cite{Gleis2023}. They are particularly important for 1-site schemes that aim to achieve lower computational costs than 2-site schemes while avoiding getting stuck in local minima when enforcing symmetry constraints.

\paragraph{Symmetries}
The ability for users to leverage good quantum numbers across a wide range of symmetry groups is an important implementation feature. 
Although enforcing these constraints introduces some overhead and can occasionally complicate convergence, the overhead is often negligible, and symmetry-adapted DMRG typically delivers significant performance gains by reducing computational cost and improving accuracy.

\paragraph{Mixed-precision}
Lower floating-point precision reduces memory usage and accelerates computations at the expense of accuracy; however, because the DMRG algorithm is iterative, the loss of accuracy is often insignificant during the initial sweeps, when the MPS is still a crude approximation.
The availability of mixed-precision strategies---such as the approach introduced by Tian et al.~\cite{Tian2022}---is therefore attractive, enabling users to perform early sweeps in reduced precision to accelerate the calculation without compromising final accuracy.

Furthermore, mixed-precision techniques that emulate double-precision arithmetic have recently shown promising results on state-of-the-art hardware \cite{Brower2026}.

\paragraph{Linear algebra libraries}
DMRG relies heavily on tensor algebra, with frequent contractions and decompositions at its core. The efficiency of these operations depends critically on the underlying linear algebra libraries. As such, selecting libraries well optimized for the target hardware is essential for achieving maximum performance.

\paragraph{Out-of-core}
Large-scale DMRG calculations---e.g., high bond dimensions or many sites---can require substantial memory. In such cases, in-memory storage may become impractical, and using out-of-core strategies for storing data on disk may be necessary to complete the computation.

\paragraph{Quantum chemistry}
In quantum chemistry applications, additional choices---such as selecting the appropriate orbital basis and ordering to minimize long-range entanglement---can significantly improve DMRG performance by reducing the effective bond dimension and accelerating convergence \cite{OlivaresAmaya2015}.

\paragraph{Parallelism}
High-performance DMRG implementations often rely on multiple levels of parallelism to accelerate computations, including GPU acceleration where supported \cite{Zhai2021, Tian2023}. Such optimizations are essential for enabling large-scale calculations to maintain a practical time-to-solution.

\section{Performance benchmarking}
\label{sec: Performance benchmarking}

\subsection{Related work}

In many areas of computational science, standardized benchmark suites provide a common framework for evaluating software performance, enabling reproducible and rigorous comparisons across implementations and hardware platforms.
For example, the LINPACK benchmark has long served as the standard for assessing high-performance computing systems and underpins the TOP500 ranking of supercomputers \cite{Dongarra2003}.
Similar benchmark suites, such as SparseBench \cite{Dongarra2001}, Graph500 \cite{Murphy2010}, HPCG \cite{Dongarra2015}, and MLPerf \cite{Mattson2020}, are widely used in their related fields where they facilitate the evaluation of algorithmic and software developments.
In contrast, performance benchmarking for DMRG implementations remains scarce and lacks a consistent standard.

While DMRG has been studied extensively from both algorithmic and application perspectives \cite{Schollwck2011, Stoudenmire2012, Chan2011, Baiardi2020}, relatively little attention has been paid to the systematic performance evaluation and benchmarking of its many implementations \cite{Sehlstedt2026}.
As such, existing benchmarks, which frequently appear in DMRG papers introducing new software packages or developments, typically serve purposes other than direct performance comparisons between existing implementations. 

For example, many software papers introducing new packages or package versions often focus on theory, applications, and usage. Consequently, a common benchmarking practice in these papers is to measure sweep time as a function of bond dimension to validate expected time-complexity scaling---as seen in papers introducing \package{ITensor} \cite{Fishman2022}, \package{Block2} \cite{Zhai2023}, \package{TeNPy} \cite{Hauschild2024}, \package{YASTN} \cite{Rams2025}, \package{Cytnx} \cite{Wu2025}, \package{pyTTN} \cite{Lindoy2025}, and \package{TensorKit} \cite{Devos2025}. Another common practice in these papers is to evaluate specific aspects, such as parallel efficiency, or to demonstrate certain capabilities, such as analysis of specific systems, which we see in the introductions of \package{REGO} \cite{Kurashige2009}, \package{ALPS MPS} \cite{Dolfi2014}, \package{CheMPS2} \cite{Wouters2014CheMPS2}, \package{MOLMPS} \cite{Brabec2020}, \package{Kylin} \cite{Song2025}, and \package{QCMaquis} \cite{Szenes2025}. Others---such as introductions of \package{DMRG++} \cite{Alvarez2009}, \package{BAGEL} \cite{Shiozaki2017}, \package{OSMPS} \cite{Jaschke2018}, \package{quimb} \cite{Gray2018}, \package{PyTeNet} \cite{Mendl2018}, \package{QSpace} \cite{Weichselbaum2024}, and \package{SeeMPS} \cite{GarciaMolina2026}---focus solely on the supported methods and features, and thus do not include any benchmarks.

Likewise, papers introducing new DMRG developments typically focus on theoretical aspects and therefore often aim to quantify the potential impact in general rather than how it compares to existing implementations. Examples include the use of non-abelian symmetries \cite{McCulloch2002, Alvarez2012, Weichselbaum2012}, density matrix perturbation \cite{White2005}, subspace expansion \cite{Hubig2015}, various parallelism strategies \cite{Zhai2021, Hager2004, Nemes2014, Kurashige2009, Solomonik2014, Chan2004, Wouters2014, Chan2016, Stoudenmire2013, Chen2021}, mixed-precision strategies \cite{Tian2022}, controlled bond expansion \cite{Gleis2023}, heterogeneous computing platforms \cite{Tian2023, Nemes2014, Li2020, Chen2020, Ganahl2023, Hong2022, Liu2024}, and emulated FP64 arithmetic \cite{Brower2026}.

In both cases, it is unsurprising that the benchmarks do not meet the requirements needed to enable meaningful performance comparisons between implementations and the needs of users seeking them, since the benchmarks have other main objectives and are often only applied to individual implementations.

While there exist a few comprehensive performance evaluations of specific packages, such as \package{tensor-tools} \cite{Levy2020}, \package{DMRG-Budapest} \cite{Menczer2025Massively, Menczer2024Tensor, Menczer2024Two, Menczer2024Parallel, Menczer2024Cost, Legeza2025}, \package{Focus} \cite{Xiang2024}, and \package{Quantum TEA} \cite{Jaschke2026}, these studies have mostly been performed in isolation, using different benchmark problems and experimental setups, and are not enough to be representative of the entire landscape of implementations. As such, evaluations at large remain incomplete, with very few cross-comparisons. This article aims to contribute toward addressing that gap.

\subsection{Requirements for performance benchmarks}

What requirements must a benchmark satisfy to enable meaningful performance comparisons between DMRG implementations? We argue that at least five fundamental conditions are necessary: (i) Computational cost and solution quality must be evaluated jointly, (ii) end-to-end performance metrics must be considered, (iii) a common set of model problems must be used, (iv) parameter configurations must be handled in a controlled manner, and (v) hardware and software environments must be handled in a controlled manner.

First, computational cost is only meaningful when considered alongside the solution quality obtained. For example, if implementation A is reported to be faster than implementation B but produces a lower-quality result, comparing time alone is not meaningful, since both implementations may ultimately require the same amount of time to reach a target quality.

Second, comparisons must be based primarily on end-to-end performance metrics rather than local metrics such as sweep times. For example, if implementation A performs sweeps faster than implementation B but requires more sweeps, sweep time alone provides an incomplete picture of end-to-end time cost and is thus not meaningful. Although local metrics remain valuable for understanding specific implementation characteristics, they need to be complemented by end-to-end metrics.

Third, the choice of model can greatly impact implementation performance. Thus, if implementation A is evaluated on model M and implementation B on model N, a comparison between them is not meaningful, since the observed difference may reflect the properties of the models rather than the implementations themselves. While all existing benchmarks evaluate all included implementations on the same set of model problems, different papers often use different benchmarks---with distinct sets of implementations and model problems---making meaningful comparisons across reported results difficult.

Fourth, the parameter configuration can significantly affect implementation performance. Therefore, handling parameter configurations in a controlled manner is necessary.

Finally, many implementations are continuously optimized and developed, and many performance metrics are inherently hardware-dependent and can vary significantly across different architectures. Therefore, handling hardware and software environments in a controlled manner is necessary.

\subsection{Performance metrics}

Having established that performance benchmarks must jointly evaluate computational cost and solution quality to enable meaningful comparisons between implementations, we need metrics that consistently and meaningfully quantify these aspects. Several metrics can be used for this purpose, each capturing different aspects of performance. Regarding metrics for solution quality, three standard ones for DMRG are:

\begin{itemize}
    \item \textbf{Energy residuals:}
    Solution quality is naturally assessed by accuracy. In DMRG, accuracy is best quantified through the computed energy and energy residuals. Because DMRG is variational and guarantees an upper bound on the ground-state energy, and because the true ground-state energy is a global minimum, lower energies indicate a better outcome of the algorithm.

    \item \textbf{Energy variance:}
    Similar to energy residuals, while not a direct indicator of accuracy, a relevant metric for solution quality is the energy variance $\sigma_H^2$ \cite{Stoudenmire2012, Hubig2018}:
    \begin{equation}
        \label{eq: energy variance}
        \sigma_H^2 = \ev{\widehat{H}^2}{\psi} - \ev{\widehat{H}}{\psi}^2.
    \end{equation}
    The closer $\sigma_H^2$ is to zero, the more accurately $\ket{\psi}$ represents an eigenstate of the Hamiltonian. However, low variance does not guarantee that the state is the ground state. 

    \item \textbf{Observables:}
    Observables other than energy can sometimes also help assess solution quality. For example, properties associated with specific symmetries---such as magnetization---can be used to check the quantum-number distribution and ensure that the state lies in the expected sector, providing complementary insight.
\end{itemize}

Regarding metrics for computational cost, three standard ones include:

\begin{itemize}
    \item \textbf{Wall time:}
    As a direct measure of time-to-solution, wall time is a practical and intuitive metric. 

    \item \textbf{Memory usage:}
    High peak memory usage can limit the feasibility of large-scale or high-accuracy DMRG simulations and can therefore be an important metric.

    \item \textbf{FLOPs:}
    The number of floating-point operations (FLOPs) performed is largely hardware-independent and can therefore facilitate comparisons across different hardware systems and provide a useful basis for analyzing implementation efficiency.
\end{itemize}

Based on measurements of these primary cost metrics, several derived performance metrics can be calculated to add further depth to the analysis. Common metrics include: 

\begin{itemize}
    \item \textbf{Floating-point throughput and efficiency:}
    Achieved floating-point throughput (FLOP/s) is commonly used to calculate the efficiency relative to the hardware's theoretical peak performance, and indicates how effectively the available computational resources are utilized.

    \item \textbf{Speedup:}
    Speedup quantifies relative speed improvements and is commonly used to evaluate the effectiveness of algorithmic optimizations and parallel implementations.

    \item \textbf{Scalability:}
    Scalability shows how efficiently the implementation uses available hardware as compute resources and problem sizes increase. This can include both strong and weak scaling across shared-memory threading, distributed parallelism, and accelerator-based execution such as GPUs.
    
    \item \textbf{I/O performance}
    For large-scale simulations, I/O can contribute significantly to the total runtime. Metrics such as I/O throughput can quantify this impact.
\end{itemize}

\section{Experimental setup and procedure}
\label{sec: Experimental setup}

\subsection{Software packages}

For our experiments, we test eight open source packages. Together, they span a diverse range of programming languages, maturity levels, and popularity, from well-established tools to more recent developments \cite{Sehlstedt2026}.

More specifically, the eight packages used in our experiments, together with the corresponding versions and a brief description of each, are:

\begin{itemize}
    \item \package{Block2} (v0.5.3)~\cite{Zhai2023, Block2SourceCode} provides a comprehensive set of DMRG algorithms for use in electronic structure methods and other applications.

    \item \package{ITensorMPS.jl} (v0.3.28)~\cite{ITensorMPS.jlSourceCode} provides high-level MPS and MPO functionality, and is part of the broader \package{ITensor}~\cite{Fishman2022} ecosystem, building on \package{ITensors.jl}~\cite{ITensors.jlSourceCode}, which offers abstract index management. 

    \item \package{MPSKit.jl} (v0.13.10)~\cite{MPSKit.jlSourceCode} provides high-level tensor network algorithms. It is part of the QuantumKitHub~\cite{QuantumKitHubGitOrg} ecosystem and builds on \package{TensorKit.jl}~\cite{Devos2025}, which supports generic symmetries.

    \item \package{PyTeNet} (v1.2)~\cite{Mendl2018, PyTeNetSourceCode} implements quantum tensor network operations and simulations structured around MPS and MPO classes. It acts as a facilitator of algorithmic experimentation.

    \item \package{quimb} (v1.12.1)~\cite{Gray2018, quimbSourceCode} is a library for quantum information and many-body calculations, focusing primarily on tensor networks.

    \item \package{Renormalizer} (v0.0.11)~\cite{Ren2018, Li2020, Jiang2020, Ren2022, RenormalizerSourceCode} is a tensor network package focused on electron-phonon quantum dynamics. 

    \item \package{TeNPy} (v1.1.0)~\cite{Hauschild2018, Hauschild2024, TeNPySourceCode} is a library for simulating strongly correlated quantum systems with tensor networks. 

    \item \package{YASTN} (v1.6.0)~\cite{Rams2025, YASTNSourceCode} is a package for differentiable linear algebra with block-sparse tensors. 
    
\end{itemize}

\subsection{Models}

For our experiments, we focus on three standard local lattice models common to condensed matter physics settings: the transverse-field Ising model, the isotropic spin-1 Heisenberg model, and the Fermi--Hubbard model. Together, these models span a range of local Hilbert-space dimensions, entanglement structures, and symmetry properties, providing a representative and challenging testbed for assessing DMRG performance.

\subsubsection{Transverse-field Ising model}

We express the transverse field Ising model with the following Hamiltonian:
\begin{equation}
    \label{eq: transverse field Ising model}
    \widehat{H} = -J \sum_{\abrc{i, j}} \sigma_i^x \sigma_j^x - h \sum_i \sigma_i^z,
\end{equation}
where $\sigma_i^x$ and $\sigma_i^z$ are Pauli operators acting on site $i$, $\abrc{i, j}$ denotes the set of unordered nearest-neighbor pairs, $J$ is the nearest-neighbor Ising exchange coupling, and $h$ is the transverse field strength. Important features of this Hamiltonian include a $\mathbb{Z}_2$ symmetry corresponding to a global spin flip and a quantum critical point separating ordered and disordered phases at $h/J=1$. The associated quantum phase transition leads to long-range correlations at criticality, providing a non-trivial setting for assessing DMRG performance.

Furthermore, for one-dimensional chains with open boundary conditions, analytical expressions for the ground-state energy are available for finite systems~\cite{Pfeuty1970}. In particular, the ground-state energy $E_0$ is given by
\begin{equation}
    E_0/J = 1 - \csc\brc{ \frac{\pi}{2(2L + 1)} },
    \label{eq: TFIM analytical energy}
\end{equation}
which provides a useful reference for benchmarking numerical results.

\subsubsection{Spin-1 Heisenberg model}

We express the isotropic spin-1 Heisenberg model with the following Hamiltonian:
\begin{equation}
    \label{eq: isotropic Heisenberg model}
    \widehat{H} = - J \sum_{\abrc{i, j}} \vec{S}_i \cdot \vec{S}_j \\
    = - J \sum_{\abrc{i, j}} \brc{\frac{1}{2} \Brc{S_i^+ S_j^- + S_i^- S_j^+} + S_i^z S_j^z},
\end{equation}
where J is the exchange coupling constant, $\vec{S}_i = (S_i^z, S_i^y, S_i^z)$ are spin-1 operators acting on site $i$, and $S_i^\pm$ are the corresponding ladder operators. It has full $\mathrm{SU}(2)$ spin-rotation symmetry corresponding to total spin conservation, with total magnetization conservation arising from its $\mathrm{U}(1)$ subgroup. The model was also among the first to which DMRG was applied~\cite{White1993}.

There are no analytical solutions for the ground state energy of the spin-1 Heisenberg antiferromagnet, as the model is non-integrable for spin $\ge 1$~\cite{Muller1987}. Nevertheless, numerical studies of finite periodic chains provide estimates in the thermodynamic limit, yielding a ground-state site energy of approximately $E_0 = -1.401 484 038 971 |J|$~\cite{WhiteHuse1993, Golinelli1994}.

\subsubsection{Fermi--Hubbard model}

We express the Fermi--Hubbard model with the following Hamiltonian:
\begin{equation}
    \label{eq: Fermi-Hubbard model}
    \widehat{H} = -t \sum_{\abrc{i,j}, \sigma} \brc{ \hat{c}_{i,\sigma}^\dagger \hat{c}_{j,\sigma} + \hat{c}_{j,\sigma}^\dagger \hat{c}_{i, \sigma} } + U \sum_i \hat{n}_{i, \uparrow} \hat{n}_{i,\downarrow}, 
    \quad 
    \hat{n}_{i,\sigma} = \hat{c}_{i,\sigma}^\dagger \hat{c}_{i, \sigma},
\end{equation}
where $t$ is the hopping amplitude, $U$ is the coupling strength, and $\hat{c}_{i,\sigma}^\dagger$ and $ \hat{c}_{i,\sigma}$ are the fermionic creation and annihilation operators of an electron of spin $\sigma \in \cbrc{\uparrow, \downarrow}$ at site $i$, respectively. Each site can be empty, occupied by a single spin-up or spin-down electron, or doubly occupied, forming a 4-dimensional local Hilbert space. The model has a rich symmetry structure, namely $\mathrm{U}(2) = \mathrm{U}(1) \times \mathrm{SU}(2)$, reflecting total particle-number and total spin conservation. Furthermore, at half-filling on bipartite lattices, the particle-number is also a pseudo-spin $\mathrm{SU}(2)$ symmetry emerging from particle-hole transformations associated with $\eta$-pairing. Together with spin $\mathrm{SU}(2)$, this enlarges the symmetry to $\mathrm{SO}(4) \cong \mathrm{SU}(2)_\eta \times \mathrm{SU}(2)_S / \mathbb{Z}_2$ \cite{Yang1989, Yang1990}.

In one dimension, the model can be solved exactly using the Bethe ansatz \cite{Lieb1968}. In particular, at half-filling in the thermodynamic limit, we get
\begin{equation}
    E_0/L = -4 t \int_0^\infty \frac{J_0(\omega) J_1(\omega)}{\omega \Brc{1 + \exp\brc{\omega U / 2 t}}}\,d\omega,
\end{equation}
where $J_0$ and $J_1$ are Bessel functions of the first kind.

\subsubsection{Model setups}

For all experiments, we use one-dimensional chains of length $L = 100$ with open boundary conditions. Regarding the model parameters, we use $h/J = 1$ for the transverse-field Ising model, \cref{eq: transverse field Ising model}, corresponding to its quantum critical point; $J = -1$ for the Heisenberg model, \cref{eq: isotropic Heisenberg model}, corresponding to the antiferromagnetic case; and $U/t = 8$ at half-filling for the Fermi--Hubbard model, \cref{eq: Fermi-Hubbard model}, representing an intermediate-coupling regime.

To validate our experimental results later, we establish reference energies for each model. For the transverse-field Ising model, we obtain the reference energy from \cref{eq: TFIM analytical energy}, giving $E_0 = -126.961876739681$. For the Heisenberg model, we compute the reference energy using \package{Block2} and \package{MPSKit} with $\mathrm{SU(2)}$ symmetry enforcement and a bond dimension of 1600, obtaining $E_0 = -138.940086166525$. Similarly, for the Fermi--Hubbard model, using the same packages, we compute the reference energy with $\mathrm{U(1)} \times \mathrm{SU(2)}$ symmetry enforcement and a bond dimension of 1600 to be $E_0 = -32.545776173096$.
These computed reference energies are consistent with their corresponding thermodynamic-limit values once finite-size effects and the use of open boundary conditions are accounted for.

\subsection{DMRG setup and measurement}

For our experiments, we initialize the MPS as a random state with a fixed bond dimension. To ensure that statistical fluctuations due to the randomness do not affect the results, we average over 10 independent runs.

We only measure performance for the DMRG call, so we exclude the time spent initializing the MPS and constructing the Hamiltonian. For Julia packages, we perform initial warm-up runs to trigger JIT compilation, followed by measurement runs. This ensures that timings reflect the implementation's performance rather than one-time compilation overhead.

We assess solution quality using the relative error in energy,
\begin{equation}
    \label{eq: relative energy error}
    \varepsilon_r = \frac{E - E_0}{\vbrc{E_0}},
\end{equation}
and computational cost using wall time. Additionally, we record these metrics at the end of every sweep to examine convergence behavior, providing a more complete picture.

The code used to generate all results presented in this work is publicly available in an online repository \cite{DMRG-BenchmarksGit}. The repository contains the source code, input files, and instructions necessary to reproduce the calculations and figures. The specific version of the code used for this study is archived and tagged to ensure reproducibility.

All calculations were performed on a single core of an Intel Xeon Gold 6132 (Skylake-SP) compute node.

\section{Results}
\label{sec: Results}

\subsection{Package comparisons}

In this section, we present and compare the performance of the software packages, including testing the symmetries each package supports, while aiming to align the simulation parameters as closely as possible across implementations. We do not claim that these settings are optimal for any particular package, nor do we employ advanced or composite strategies. Instead, we use a set of simple, fixed parameter settings across all sweeps to obtain a baseline performance.

\begin{figure}[htb]
    \centering
    \includegraphics{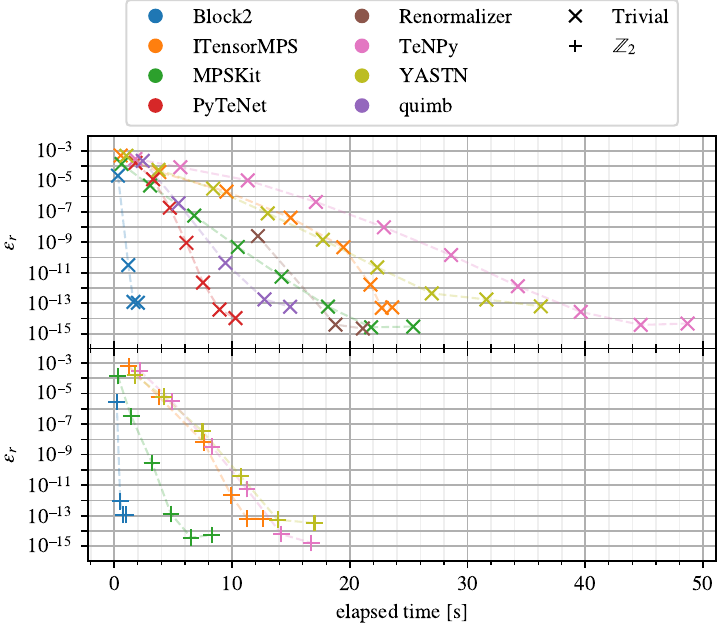}
    \caption{Performance of DMRG implementations on the transverse-field Ising model using a maximum bond dimension of 100. The top and bottom panels show trivial and $\mathbb{Z}_2$ symmetry enforcement, respectively. Markers mark the end of a full sweep, i.e., a forward and backward pass.}
    \label{fig: I100-package-symmetry}
\end{figure}

\begin{figure}[htb]
    \centering
    \includegraphics{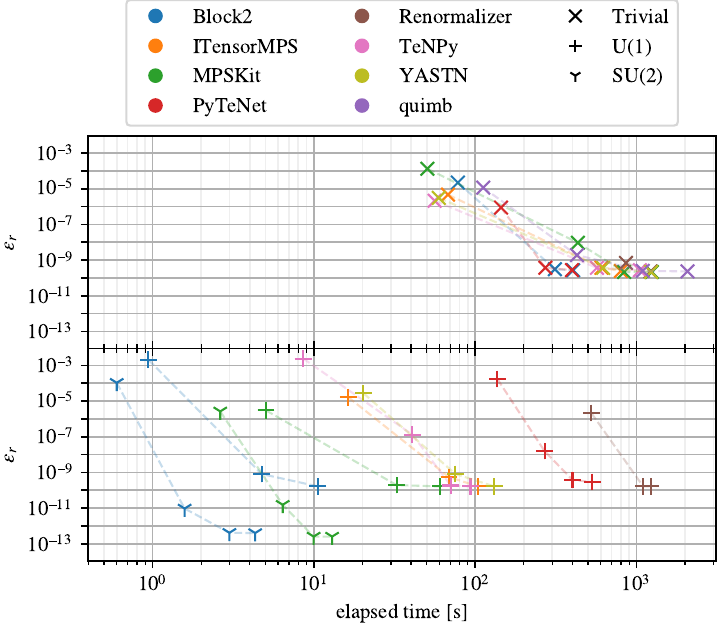}
    \caption{Performance of DMRG implementations on the isotropic spin-1 Heisenberg model using a maximum bond dimension of 400. The top and bottom panels show trivial and non-trivial symmetry enforcement, respectively. Markers mark the end of a full sweep, i.e., a forward and backward pass.}
    \label{fig: H400-package-symmetry}
\end{figure}

\begin{figure}[htb]
    \centering
    \includegraphics{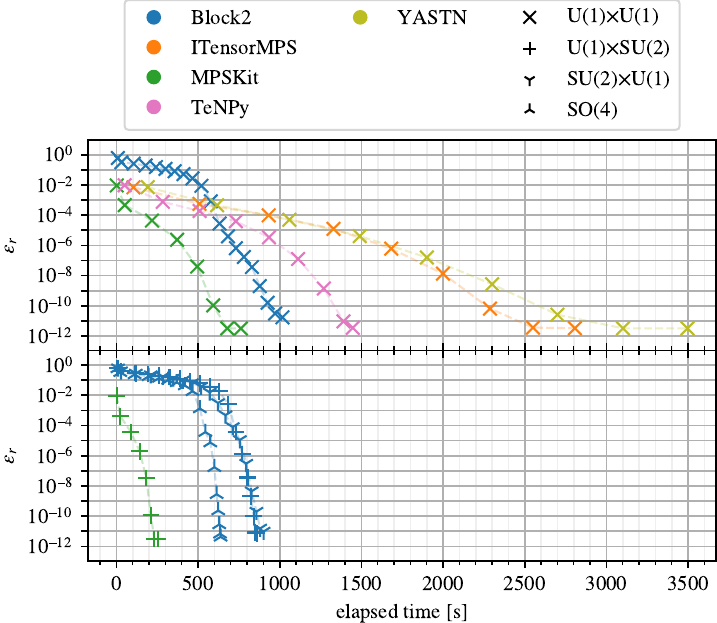}
    \caption{Performance of DMRG implementations on the Fermi--Hubbard model using a maximum bond dimension of 800. The top and bottom panels show the enforcement of abelian and non-abelian symmetries, respectively. Markers mark the end of a full sweep, i.e., a forward and backward pass.}
    \label{fig: FH800-package-symmetry}
\end{figure}

We show the results for the transverse-field Ising model in \Cref{fig: I100-package-symmetry}, the isotropic spin-1 Heisenberg model in \Cref{fig: H400-package-symmetry}, and the Fermi--Hubbard model in \Cref{fig: FH800-package-symmetry}. In the Fermi--Hubbard test, not all software packages are included: \package{quimb} does not support the required symmetry, \package{PyTeNet} converged to a different energy of -232.545, and \package{Renormalizer} did not achieve better than $10^{-2}$ accuracy after 18 sweeps and multiple hours of computation, with the given settings. Nevertheless, across all three models, the results provide a rich basis for comparison.

We observe considerable variation in performance across the packages examined, sometimes with multiple orders of difference in elapsed time to reach a given accuracy---even when enforcing the same symmetry type. For the transverse-field Ising results in \Cref{fig: I100-package-symmetry}, for example, when enforcing no symmetry, we observe that the package reaching an accuracy of nearly $10^{-13}$ most quickly does so in less than 2~s, whereas the slowest takes nearly 40~s to cross that level, corresponding to nearly a 20-fold difference. Similarly, for the much larger Fermi--Hubbard system results in \Cref{fig: FH800-package-symmetry}, when enforcing a $\mathrm{U(1) \times \mathrm{U(1)}}$ symmetry, the package crossing an accuracy of $10^{-11}$ most quickly does so in less than 700~s, whereas the slowest takes more than 3000~s, corresponding to more than a 4-fold difference.

We also observe that including symmetries clearly impacts performance. In \Cref{fig: I100-package-symmetry}, for example, simulations enforcing $\mathbb{Z}_2$ symmetry consistently outperform those performed without symmetry enforcement. Moreover, when comparing simulations from the same package, enforcing symmetry yields a 2- to 4-fold speedup. In \Cref{fig: H400-package-symmetry}, the difference is even more pronounced, with the fastest simulations exploiting non-abelian $\mathrm{SU(2)}$ symmetry outperforming most simulations using abelian $\mathrm{U(1)}$ symmetry by nearly an order of magnitude. At the same time, the latter themselves tend to be nearly an order of magnitude faster than simulations without symmetry enforcement.

Beyond symmetries, we observe the underlying physical model influencing the relative performance rankings. Different packages carry distinct levels of computational overhead, which vary in prominence depending on the model's characteristics. Thus, testing a single model is insufficient to characterize a package's overall performance; a broader set of models is necessary to obtain a more representative and balanced comparison.

These results also show that time per sweep alone is an insufficient metric for assessing performance. For instance, comparing the $\mathrm{U(1)} \times \mathrm{U(1)}$ results of \package{Block2} and \package{MPSKit} in \Cref{fig: FH800-package-symmetry} reveals that while \package{Block2} performs sweeps faster with the given settings, \package{MPSKit} gains more accuracy faster. More specifically, \package{Block2} achieves an accuracy of nearly $10^{-10}$ in roughly 900~s by performing 18 sweeps; meanwhile, \package{MPSKit} reaches the same level of accuracy in roughly 600~s but performs only 6 sweeps. Furthermore, we observe multiple packages having shorter sweep times as they approach convergence, despite using the same settings for all sweeps; for instance, in the trivial results of \package{Renormalizer} in \Cref{fig: I100-package-symmetry}, the second sweep takes roughly 6~s, whereas the third only takes roughly 3~s. Thus, we show why it is essential to consider the end-to-end interplay between computational cost and solution quality to enable meaningful comparisons across implementations.

Finally, we cannot establish a definitive performance ranking based on these results, as these tests are confined to a specific setting that does not exploit all specialized features or targeted optimizations, which can be unique to individual implementations and could significantly boost performance. Nevertheless, the substantial variation across the packages when evaluating different models and symmetries offers valuable insights for developers and users and provides a useful starting point for further investigation.

\subsection{Parameter tuning and optimization strategies}

In this section, we test various settings and strategies for tunable parameters and implementation-specific features, and examine how these choices can affect performance.

\subsubsection{Krylov subspace dimension}

\begin{figure}[htb]
    \centering
    \includegraphics{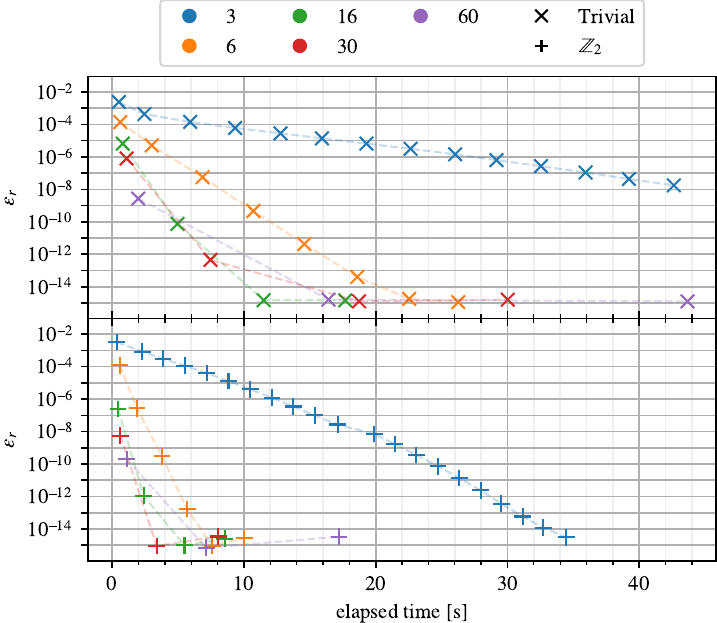}
    \caption{Performance of \package{MPSKit} on the transverse-field Ising model using different Krylov subspace dimensions.}
    \label{fig: I100-krylov}
\end{figure}

In \Cref{fig: I100-krylov}, we show how changing the Krylov subspace dimension can affect the interplay between computational cost and solution quality. 

We observe considerable variation in performance across the different settings. Using a larger Krylov subspace dimension can enable faster convergence, but setting it too large can make individual sweeps prohibitively expensive, reducing overall efficiency. As such, we again see that sweep time alone provides limited insight into performance: a faster sweep setting may still require substantially more sweeps to achieve the same accuracy, underscoring the importance of considering the end-to-end interplay between computational cost and solution quality.

Moreover, we observe that parameter choices are not independent. Here specifically, we see how different symmetry settings favor different Krylov settings. More broadly, different models will likely exhibit different optima; here, we study a critical point that will exhibit characteristics distinct from those of non-critical points. This difference suggests that there is no universally best parameter choice and no guarantee that a package's defaults are always optimal, since they are typically set to fixed values. 

Taken together, these results highlight the importance of choosing algorithmic parameters appropriately for the problem at hand to achieve optimal performance. While exploiting symmetries can substantially improve performance, these benefits are only fully realized when the other settings are chosen appropriately as well. In some cases, a poorly tuned configuration with symmetries may even perform worse than a well-tuned one without them. Thus, the practical performance of any package depends heavily on how effectively its parameters can be adapted to the problem at hand. A package having sufficient flexibility to tune and dynamically adjust parameters is therefore essential for achieving robust and efficient performance across a wide range of problems.

\subsubsection{Mixed-precision}

\begin{figure}[htb]
    \centering
    \includegraphics{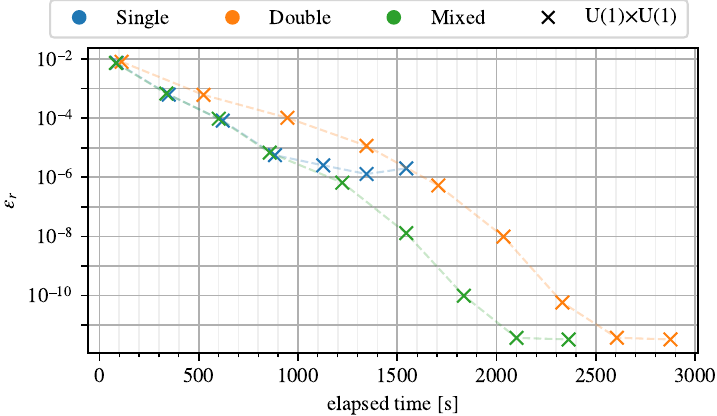}
    \caption{Performance of \package{ITensorMPS} on the Fermi--Hubbard model using three different floating-point precision strategies: single-precision only, double-precision only, and starting with single- and switching to double-precision after four sweeps. Because the single-precision Hamiltonian yields lower energy due to numerical imprecision, the single-precision result error is calculated as an absolute difference in \cref{eq: relative energy error} rather than a linear one.}
    \label{fig: FH800-mixed-precision}
\end{figure}

In \Cref{fig: FH800-mixed-precision}, we show how reduced floating-point precision can accelerate the early optimization stage without sacrificing final accuracy.

When comparing the single-precision convergence with the double-precision convergence up to an accuracy of roughly $10^{-5}$, the accuracy gain per sweep is roughly the same, but, as expected, the sweep time for the single-precision calculation is faster---single-precision requiring 900~s and double-precision requiring 1350~s, roughly, to perform four sweeps. However, beyond this point, the single-precision simulation's accuracy stagnates due to limited numerical precision, while the double-precision simulation continues to improve, reaching an accuracy below $10^{-11}$. 

This trade-off naturally suggests a mixed-precision strategy. Hence, we also show that combining the two approaches yields a clear practical advantage: By first running in single precision during the early, less sensitive stages of the optimization and then switching to double precision once higher accuracy becomes necessary, we retain the fast initial convergence while still reaching the same final accuracy as the full double-precision run. In this way, while the double-precision simulation requires nearly 2600~s to achieve an accuracy of $10^{-11}$, the hybrid method requires only 2100~s, corresponding to nearly a 1.24× speedup.

Despite the practical advantage, few packages support mixed-precision workflows directly \cite{Sehlstedt2026}. Although users can, in many cases, achieve similar behavior by manually restarting or switching precision during optimization, such interventions are cumbersome and are therefore often avoided in practice.

More broadly, these results illustrate the importance of remembering that the DMRG algorithm is iterative, and the optimal settings may change during different stages of an optimization. Hence, flexibility can be essential for achieving maximal performance.

\subsubsection{Update mode}

\begin{figure}[htb]
    \centering
    \includegraphics{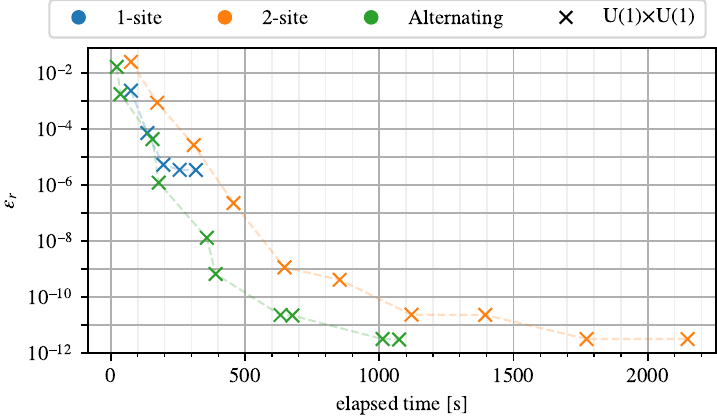}
    \caption{Performance of \package{YASTN} on the Fermi-Hubbard model using three update mode strategies: 1-site only, 2-site only, and alternating 1-site and 2-site.}
    \label{fig: FH800-mode}
\end{figure}

In \Cref{fig: FH800-mode}, we show how using different update modes---either 1-site or 2-site---at different stages can improve performance by leveraging their respective strengths and minimizing their drawbacks.

We observe that the 1-site method performs sweeps faster for a given bond dimension because it is more computationally cost-effective than the 2-site method. Still, it becomes stuck in a suboptimal configuration, as it tends to do when symmetries are enforced, and it reaches only an accuracy of $10^{-5}$. Moreover, since the traditional 1-site method cannot dynamically increase the bond dimension, it requires starting with the target bond dimension, which can be inefficient during the early stages of the optimization when a large bond dimension is not yet necessary. In contrast, the 2-site method, while slower in sweep times, is more robust and can offset part of the overhead through its adaptability: starting with a small bond dimension and gradually increasing it allows the method to focus computational effort where it is most needed.

Building on these complementary strengths, we show that, since DMRG is iterative, we can combine the two approaches by alternating between 1-site and 2-site sweeps, starting with a small bond dimension and increasing it after every other sweep. This strategy aims to improve performance by using the efficiency of the 1-site updates while relying on the 2-site method to enhance stability and enable controlled growth of the bond dimension. As a result, the hybrid method achieves an accuracy of $10^{-11}$ in just over 1000~s, whereas the pure 2-site method takes nearly 1800~s to reach the same accuracy.

Finally, comparing these \package{YASTN} results to those in \Cref{fig: FH800-package-symmetry}, where reaching an accuracy of $10^{-11}$ required around 3100~s, simply using a more gradual increase in bond dimension resulted in nearly a 2-fold speedup, and adding the alternating 1-site and 2-site sweeps on top thus yielded more than a 3-fold speedup.

Again, these results illustrate how substantially the performance can depend on the optimization strategy and parameter choices, highlighting the importance of flexibility. While many packages provide both 1-site and 2-site optimization schemes, fewer support seamless switching between them during a calculation.

\subsubsection{Subspace expansion}

\begin{figure}[htb]
    \centering
    \includegraphics{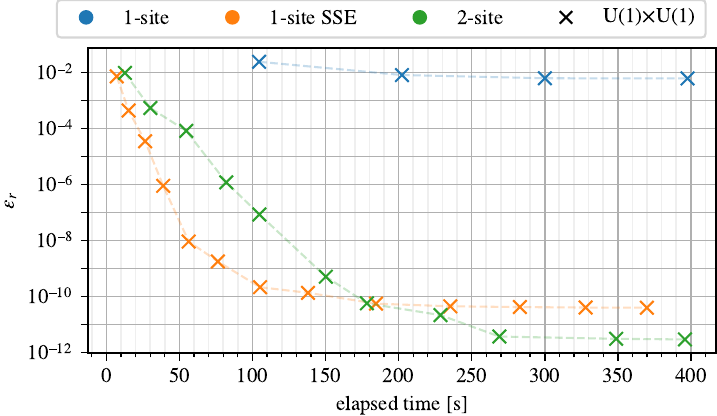}
    \caption{Performance of \package{TeNPy} on the Fermi-Hubbard model using 2-site DMRG and 1-site DMRG with and without subspace expansion.}
    \label{fig: FH800-subspace-expansion}
\end{figure}

In \Cref{fig: FH800-subspace-expansion}, we show how subspace expansion (SSE) can be used with 1-site DMRG to speed up calculations; similar to density matrix perturbation, it can help 1-site DMRG avoid getting trapped in local minima, but it also enables controlled growth of the bond dimension.

To start, we again observe the typical drawbacks of the traditional 1-site method, i.e., the need to start at the target bond dimension and the tendency to get stuck in local minima when enforcing symmetries, in this case, resulting in only $10^{-2}$ accuracy after approximately 200~s. In contrast, the SSE approach largely avoids both issues, achieving an accuracy of nearly $10^{-10}$ in just under 140~s.

Moreover, we observe that the SSE approach offers a competitive advantage over the 2-site method at the outset, as it also allows controlled growth of the bond dimension. However, in this case, after reaching an accuracy of approximately $10^{-8}$, the SSE convergence begins to slow and eventually stagnates, reaching a final accuracy just below $10^{-10}$, whereas the 2-site method converges to a final accuracy of around $5 \times 10^{-12}$. Still, up to that point, SSE provides a substantial early-time speedup; for example, we see it reach an accuracy of $10^{-6}$ in roughly 40~s, while the 2-site method requires about 80~s to achieve the same level, corresponding to a 2-fold speedup.

Furthermore, as with previous examples, DMRG is iterative, so the best strategy may be to combine the two approaches, using SSE in the low-to-moderate-accuracy regime and then switching to 2-site for the high-accuracy regime. Thus, by looking beyond the final error values themselves, we see that studying convergence behavior provides valuable insight into how each approach operates throughout the optimization process. In particular, we see how it can reveal regimes in which one approach reaches intermediate accuracies substantially faster than another, and in which the latter becomes more efficient near the precision limit. More broadly, this illustrates the value of doing more systematic testing, as it can reveal non-obvious performance trade-offs.

SSE workflows, similar to mixed-precision and mode-hybrid approaches, are supported by only a limited number of packages \cite{Sehlstedt2026}. This reflects a broader issue in the software landscape, where many packages are developed largely independently, with limited modularity and interoperability. As a result, promising techniques like SSE experience slower dissemination and adoption, despite providing practical benefits.

Finally, comparing these \package{TeNPy} results to those in \Cref{fig: FH800-package-symmetry}, where achieving an accuracy of $10^{-11}$ took nearly 1400~s, using the 2-site approach with a gradual increase in bond dimension does it in about 270~s, again highlighting just how sensitive these packages can be to the optimization strategies and parameter choices. More importantly, this suggests that selecting an effective optimization strategy can have a greater impact on performance than the choice of package itself, reinforcing the importance of systematic evaluation when assessing and comparing packages.

\section{Discussion and conclusion}
\label{sec: Discussion}

We have examined a range of packages and parameter configurations, applying the performance-oriented benchmarking framework presented in \Cref{sec: Performance benchmarking}. The results showed significant performance differences, up to two orders of magnitude in some cases. Not only were the differences between packages significant when aligning parameters, but they were also significant within the same package when comparing different parameter settings and optimization strategies.

The observed performance sensitivity to the optimization strategy raises an important methodological consideration: whether parameter settings should be aligned across packages or each package should be evaluated under conditions that allow it to fully exploit its capabilities. While aiming to align parameters may improve comparability in some sense, it may underrepresent the strengths of packages that incorporate advanced or specialized features. This issue is particularly relevant given that many of the more recent innovations---e.g., subspace expansion or mixed-precision approaches---are implemented in only a subset of packages, and even fewer packages support multiple such innovations. Furthermore, if the comparison is expanded to include many packages, the set of features and parameter settings shared across all implementations becomes increasingly limited. As a result, comparisons based solely on harmonized settings may not fully reflect the practical performance achievable by state-of-the-art implementations, hindering our ability to assess how each package truly performs relative to the others in practice.

However, attempting to evaluate packages under conditions that reflect their full potential would introduce its own set of challenges and methodological questions. Determining the optimal configuration for your own package, let alone one developed by others, for a given problem is challenging, if not impossible, due to the level of expertise required and the endless number of possible configurations. This challenge thus underscores the value of reproducibility and transparency in benchmarking studies. Even with suboptimal configurations, by clearly documenting parameter settings, optimization strategies, and simulation procedures, researchers enable others to understand, replicate, and critically assess the reported results, thereby strengthening the reliability and interpretability of the performance assessment.

The observed influence of parameter configuration on performance also motivates broader questions regarding the role of parameter tuning itself. In particular, it remains unclear to what extent effective performance requires extensive manual tuning, or whether more general and automated tuning procedures could achieve comparable results across a range of problems and packages.

More generally, our findings raise questions about whether certain parameters have a disproportionately large impact on performance across a wide range of problems, while others play a comparatively minor role. For example, for the models we studied, we observe multiple parameter settings that can significantly impact performance. Moreover, a similar related question is the extent to which optimal parameter settings and optimization strategies depend on the specific problem under consideration; whether some require substantial adjustment across different problem classes, whereas others exhibit robust near-optimal settings that generalize well across a broad range. Identifying such influential parameters and strategies could help focus tuning efforts and simplify practical use. 

The results thus also highlight the importance of understanding interactions between parameters and optimization strategies. In particular, it remains unclear whether we can further combine certain strategies to achieve additional performance gains or whether their effects would overlap and interfere negatively with one another. Addressing these questions requires that packages support all available strategies and provide sufficient flexibility to combine them seamlessly. Such flexibility will likely also be essential for achieving maximal performance across a wide range of problems.

Taken together, one thing is for certain: There is significant value and insight to be gained from conducting more rigorous performance evaluations. When they are performed, there needs to be consideration of the interplay between computational cost and solution quality to enable meaningful comparisons. Using the presented benchmarking framework and our results as a starting point, continued benchmarking will ultimately help users and developers make informed decisions and support future development efforts to build better, more efficient software.

\printbibliography

\end{document}